\documentclass[twocolumn]{article}  
\usepackage[left=2.5cm, right=2.5cm,top=2.5cm,bottom=3cm]{geometry}
\usepackage{graphicx}
\usepackage{mathtools}
\usepackage{color}
\usepackage{upgreek}
\usepackage{float}
\usepackage{xspace}
\usepackage{booktabs} 
\usepackage{txfonts}
\usepackage[comma,super,sort&compress]{natbib}
\usepackage[colorlinks = true,
            linkcolor = blue,
            urlcolor  = blue,
            citecolor = blue,
            anchorcolor = blue]{hyperref}
\usepackage[noblocks]{authblk}

\newcommand\arcmin{\mbox{$^\prime$}}%
\newcommand\arcsec{\mbox{$^{\prime\prime}$}}%
\newcommand\farcs{\mbox{$.\!\!^{\prime\prime}$}}%
\newcommand{\apjs}{Astrophys. J. Suppl. Ser.} 
                                 
\newcommand{\aap}{Astron. Astrophys.} 
                                
\newcommand{\apjl}{Astrophys. J. Lett.} 

\newcommand{\rone}{FRB~20121102A\xspace}
\newcommand{\rthree}{FRB~20180916B\xspace}
\newcommand{\frb}{FRB~20200120E\xspace}

\newcommand{\gc}{[PR95]~30244\xspace}
\newcommand{\rsixseven}{FRB~20201124A\xspace}

\newcommand{\dmunit}{pc\,cm$^{-3}$\xspace}
\newcommand{\us}{$\upmu$s\xspace}

\makeatother

\begin{document} 

\begin{titlepage}
\newgeometry{top=1cm, bottom=2cm, left=2cm, right=2cm}
\onecolumn
\title{A repeating fast radio burst source in a globular cluster}
\author[1]{F.~Kirsten}
\author[2]{B.~Marcote}
\author[3,4]{K.~Nimmo}
\author[4,3]{J.~W.~T.~Hessels}
\author[5,6]{M.~Bhardwaj}
\author[7,8]{S.~P.~Tendulkar}
\author[2]{A.~Keimpema}
\author[1]{J.~Yang}
\author[4]{M.~P.~Snelders}
\author[9]{P.~Scholz}
\author[5,6,10,11,12]{A.~B.~Pearlman}
\author[13,14]{C.~J.~Law}
\author[15]{W.~M.~Peters}
\author[16]{M.~Giroletti}
\author[2]{Z.~Paragi}
\author[3]{C.~Bassa}
\author[4]{D.~M.~Hewitt}
\author[17]{U.~Bach}
\author[18]{V.~Bezrukovs}
\author[19]{M.~Burgay}
\author[20]{S.~T.~Buttaccio}
\author[1]{J.~E.~Conway}
\author[19]{A.~Corongiu}
\author[21]{R.~Feiler}
\author[1]{O.~Forssén}
\author[21]{M.~P.~Gawroński}
\author[17]{R.~Karuppusamy}
\author[22]{M.~A.~Kharinov}
\author[1]{M.~Lindqvist}
\author[16]{G.~Maccaferri}
\author[22]{A.~Melnikov}
\author[4]{O.~S.~Ould-Boukattine}
\author[19,23]{A.~Possenti}
\author[19]{G.~Surcis}
\author[24]{N.~Wang}
\author[24]{J.~Yuan}
\author[25,26]{K.~Aggarwal}
\author[25,26]{R.~Anna-Thomas}
\author[27]{G.~C.~Bower}
\author[3]{R.~Blaauw}
\author[25,26,28]{S.~Burke-Spolaor}
\author[9,29]{T.~Cassanelli}
\author[15]{T.~E.~Clarke}
\author[5,6,25,26]{E.~Fonseca}
\author[9,29]{B.~M.~Gaensler}
\author[4]{A.~Gopinath}
\author[5,6]{V.~M.~Kaspi}
\author[15]{N.~Kassim}
\author[30]{T.~J.~W.~Lazio}
\author[31,32]{C.~Leung}
\author[13]{D.~Z.~Li}
\author[33]{H.~H.~Lin}
\author[31,32]{K.~W.~Masui}
\author[9]{R.~Mckinven}
\author[5,6]{D.~Michilli}
\author[22]{A.~Mikhailov}
\author[9]{C.~Ng}
\author[18]{A.~Orbidans}
\author[33,9,28,34]{U.~L.~Pen}
\author[4,5,6,35]{E.~Petroff}
\author[36]{M.~Rahman}
\author[37]{S.~M.~Ransom}
\author[31,32]{K.~Shin}
\author[34]{K.~M.~Smith}
\author[38]{I.~H.~Stairs}
\author[1]{W.~Vlemmings}
\affil[1]{Department of Space, Earth and Environment, Chalmers University of Technology, Onsala Space Observatory, 439 92, Onsala, Sweden}
\affil[2]{Joint Institute for VLBI ERIC, Oude Hoogeveensedijk 4, 7991 PD Dwingeloo, The Netherlands}
\affil[3]{ASTRON, Netherlands Institute for Radio Astronomy, Oude Hoogeveensedijk 4, 7991 PD Dwingeloo, The Netherlands}
\affil[4]{Anton Pannekoek Institute for Astronomy, University of Amsterdam, Science Park 904, 1098 XH, Amsterdam, The Netherlands}
\affil[5]{Department of Physics, McGill University, 3600 rue University, Montréal, QC H3A 2T8, Canada}
\affil[6]{McGill Space Institute, McGill University, 3550 rue University, Montréal, QC H3A 2A7, Canada}
\affil[7]{Department of Astronomy and Astrophysics, Tata Institute of Fundamental Research, Mumbai, 400005, India}
\affil[8]{National Centre for Radio Astrophysics, Post Bag 3, Ganeshkhind, Pune, 411007, India}
\affil[9]{Dunlap Institute for Astronomy \& Astrophysics, University of Toronto, 50 St. George Street, Toronto, ON M5S 3H4, Canada}
\affil[10]{Division of Physics, Mathematics, and Astronomy, California Institute of Technology, Pasadena, CA 91125, USA}
\affil[11]{McGill Space Institute~(MSI) Fellow}
\affil[12]{FRQNT Postdoctoral Fellow}
\affil[13]{Cahill Center for Astronomy and Astrophysics, California Institute of Technology, 1216 E California Boulevard, Pasadena, CA 91125, USA}
\affil[14]{Owens Valley Radio Observatory, California Institute of Technology, Pasadena, CA 91125, USA}
\affil[15]{Remote Sensing Division, US Naval Research Laboratory, Washington, DC 20375, USA}
\affil[16]{Istituto Nazionale di Astrofisica, Istituto di Radioastronomia, via Gobetti 101, I-40129 Bologna, Italy}
\affil[17]{Max Planck Institute for Radio Astronomy, Auf dem Heugel 69 53121 Bonn, Germany}
\affil[18]{Engineering Research Institute Ventspils International Radio Astronomy Centre (ERI VIRAC) of Ventspils University of Applied Sciences (VUAS), Inzenieru street 101, Ventspils, LV-3601, Latvia}
\clearpage
\affil[19]{Istituto Nazionale di Astrofisica, Osservatorio Astronomico di Cagliari, Via della Scienza 5, I-09047, Selargius, Italy}
\affil[20]{Istituto Nazionale di Astrofisica, Istituto di Radioastronomia Radiotelescopio di Noto, C.da Renna Bassa Loc. Case di Mezzo C.P. 161, 96017, Noto (SR), Italy}
\affil[21]{Institute of Astronomy, Faculty of Physics, Astronomy and Informatics, Nicolaus Copernicus University, Grudziadzka 5, 87-100 Toru'n, Poland}
\affil[22]{Institute of Applied Astronomy of the Russian Academy of Sciences, Kutuzova Embankment 10, St. Petersburg, 191187, Russia}
\affil[23]{Università di Cagliari, Dipartimento di Fisica, S.P. Monserrato-Sestu Km 0,700 - 09042 Monserrato (CA), Italy}
\affil[24]{Xinjiang Astronomical Observatory, 150 Science 1-Street, Urumqi, Xinjiang 830011, China}
\affil[25]{Department of Physics and Astronomy, West Virginia University, P.O. Box 6315, Morgantown, WV 26506, USA}
\affil[26]{Center for Gravitational Waves and Cosmology, West Virginia University, Chestnut Ridge Research Building, Morgantown, WV 26505, USA}
\affil[27]{Academia Sinica Institute of Astronomy and Astrophysics, 645 N. A'ohoku Place, Hilo, HI 96720, USA}
\affil[28]{Canadian Institute for Advanced Research, CIFAR Azrieli Global Scholar, MaRS Centre West Tower, 661 University Ave. Suite 505, Toronto ON M5G 1M1, Canada}
\affil[29]{David A. Dunlap Department of Astronomy \& Astrophysics, University of Toronto, 50 St. George Street, Toronto, ON M5S 3H4, Canada}
\affil[30]{Jet Propulsion Laboratory, California Institute of Technology, 4800 Oak Grove Dr, Pasadena, CA 91109, USA}
\affil[31]{MIT Kavli Institute for Astrophysics and Space Research, Massachusetts Institute of Technology, 77 Massachusetts Ave, Cambridge, MA 02139, USA}
\affil[32]{Department of Physics, Massachusetts Institute of Technology, 77 Massachusetts Ave, Cambridge, MA 02139, USA}
\affil[33]{Canadian Institute for Theoretical Astrophysics, University of Toronto, 60 St. George Street, Toronto, ON, M5S 3H8, Canada}
\affil[34]{Perimeter Institute for Theoretical Physics, 31 Caroline Street North, Waterloo, ON, N2L 2Y5, Canada}
\affil[35]{Veni Fellow}
\affil[36]{Sidrat Research, PO Box 73527 RPO Wychwood, Toronto, ON M6C 4A7, Canada}
\affil[37]{National Radio Astronomy Observatory, 520 Edgemont Rd., Charlottesville, VA 22903, USA}
\affil[38]{Dept. of Physics and Astronomy, University of British Columbia, 6224 Agricultural Road, Vancouver, BC V6T 1Z1}
\date{}
\maketitle
\thispagestyle{empty}
\end{titlepage}

\restoregeometry
\twocolumn
\textbf{Fast radio bursts (FRBs) are exceptionally luminous flashes of unknown physical origin, reaching us from other galaxies \citep{petroff_2019_aarv}. Most FRBs have only ever been seen once, while others flash repeatedly, though sporadically \citep{spitler_2016_natur, chime_2021_arXiv}.  Many models invoke magnetically powered neutron stars (magnetars) as the engines producing FRB emission\citep{margalit_2018_apjl, chime_2020_natur_587}. Recently, CHIME/FRB announced the discovery\citep{bhardwaj_2021_apjl} of the repeating \frb, coming from the direction of the nearby grand design spiral galaxy M81.  Four potential counterparts at other observing wavelengths were identified\citep{bhardwaj_2021_apjl} but no definitive association with these sources, or M81, could be made.  Here we report an extremely precise localisation of \frb, which allows us to associate it with a globular cluster (GC) in the M81 galactic system and to place it $\approx$2\,pc offset from optical center of light of the GC. This confirms\citep{bhardwaj_2021_apjl} that \frb is 40 times closer than any other known extragalactic FRB. Because such GCs host old stellar populations, this association strongly challenges FRB models that invoke young magnetars formed in a core-collapse supernova as powering FRB emission. 
We propose, instead, that \frb is a highly magnetised neutron star formed via either accretion-induced collapse of a white dwarf or via merger of compact stars in a binary system\citep{margalit_2019_apj}. Alternative scenarios involving compact binary systems, efficiently formed inside globular clusters, could also be responsible for the observed bursts.
}

M81 is one of the most massive nearby galaxies, at 3.6\,Mpc\citep{freedman_1994_apj}. We targeted the previously best-known position\citep{bhardwaj_2021_apjl} of \frb\ several times during February to May 2021 with an array composed of up to eleven radio telescopes that are part of the European Very Long Baseline Interferometry (VLBI) Network (EVN; see Methods). We observed at a central radio frequency of $\sim1.4$~GHz and recorded raw voltages from all telescopes. At Effelsberg and at the Sardinia Radio Telescope (SRT), we collected data in parallel using pulsar data recorders (Methods).

\begin{figure*}
\resizebox{\hsize}{!}
        {\includegraphics[trim=0cm 0cm 0cm 0cm, clip=true]{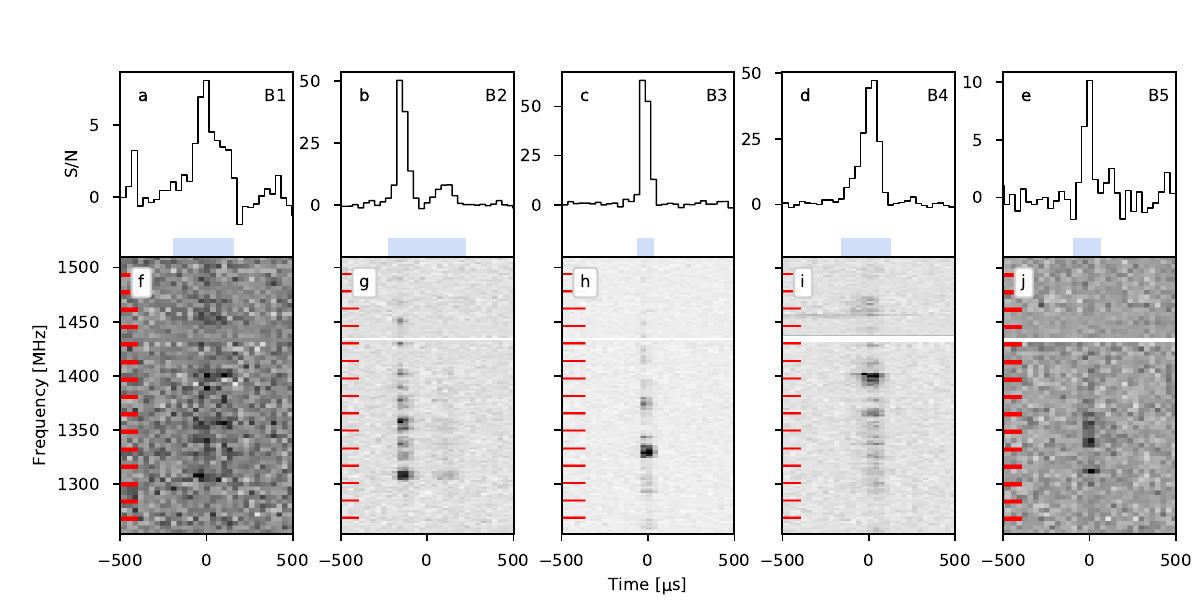}}
  \caption{\textbf{Dispersion-corrected time series and dynamic spectra of the five \frb bursts} Frequency-averaged time series of the bursts are displayed in panels \textbf{a-e}. The top right corner of each panel shows the burst name used in this work. The blue-coloured bars highlight the $\pm$2$\sigma$ burst width used to measure the burst fluence. The dynamic spectra of each burst are shown in panels \textbf{f-j}. The red marks represent the edges of the subbands. Data that have been removed due to contamination by radio frequency interference have not been plotted. In all panels the data are plotted with 32\,$\upmu$s and 2\,MHz time and frequency resolution, respectively (with the exception of B1 and B5 which are plotted with 4\,MHz frequency resolution).}
     \label{fig:bursts}
\end{figure*}

In total we detected five bursts from \frb, with dispersion measures close to the previously reported\citep{bhardwaj_2021_apjl} $\mathrm{DM=87.8}$\,\dmunit.  Two bursts were detected on 2021 February 20 (called B1 and B2 below), two bursts on 2021 March 7 (B3 and B4), and one burst on 2021 April 28 (B5). Bursts B2-B5 were found by blindly searching both the Effelsberg voltages and PSRIX\citep{lazarus_2016_mnras} data (Methods), while B1 was only detected in the voltage data as it occurred outside the recording times of the PSRIX instrument\citep{nimmo_2021_arxiv}. In Figure~\ref{fig:bursts} we show the dedispersed dynamic spectra and frequency averaged burst profiles. Burst fluences range from $0.13-0.71$\,Jy~ms and total burst envelopes span only $\sim100-300$\,\us (Table~\ref{tab:pol_properties}). A detailed, ultra-high-time-resolution analysis of the burst properties and energetics is presented in a companion paper\citep{nimmo_2021_arxiv}.

Correlation of the data, in order to produce `visibilities' for interferometric imaging, was performed at the Joint Institute for VLBI ERIC. To achieve the best-possible sensitivity, we used the coherent-dedispersion mode of the software correlator SFXC\citep{keimpema_2015_exa}, applying $\mathrm{DM=87.77\,pc\,cm^{-3}}$, which we derived from a manual inspection of the bursts (Methods). 

\begin{figure*}
        \centering
        \includegraphics[width=\textwidth]{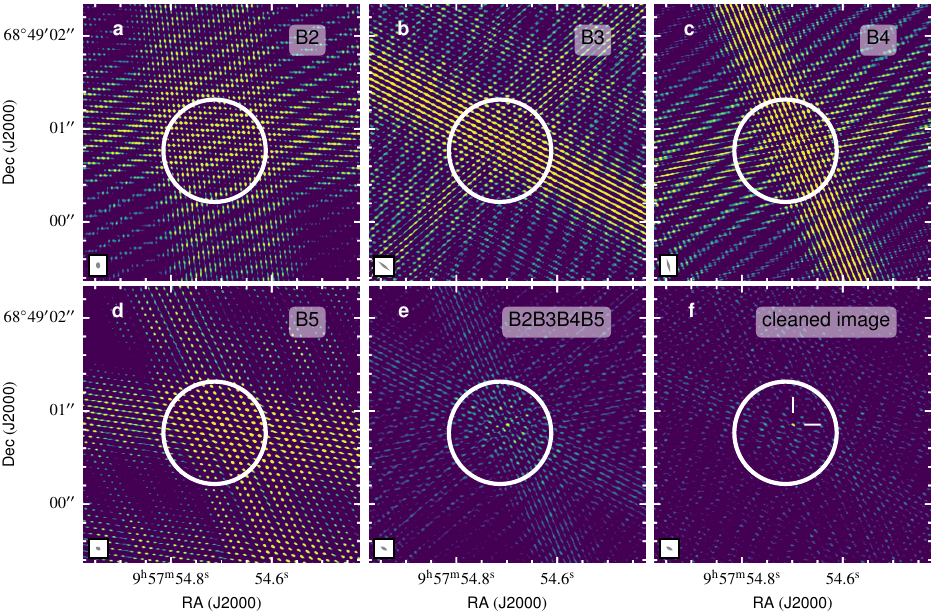}
  \caption{\label{fig:localisation-maps}\textbf{Localisation plots for \frb.} Normalised dirty images of the individual bursts (B2, B3, B4 and B5; \textbf{a-d}) along with a dirty image of the four bursts combined (\textbf{e}), produced by applying a natural weighting to the data. For visualisation purposes we clip the colour scale at zero, i.e. only positive peaks are displayed. The white circles are centred on the location of the globular cluster \gc as derived from the Subaru image in Figure~\ref{fig:subaru-image}. Their size indicates the region that contains 90\% of the globular cluster's optical emission.
  The synthesized beams of each image are displayed as grey ellipses in the bottom left corner of each panel.
  \textbf{f}: The cleaned image and resulting localisation of \frb, as derived from the combined data sets of four bursts. The resulting coordinates of \frb (highlighted by the white marker) are RA (J2000) $= 9^{\rm h}57^{\rm m}54.69935^{\rm s} \pm 1.2\,{\rm mas}$,\ Dec (J2000) $= 68^\circ49^\prime0.8529^{\prime\prime} \pm 1.3\,{\rm mas}$.}
\end{figure*}

After an initial rough localisation via delay mapping (Methods, accurate to several arcseconds), we individually imaged the five bursts, where each data set spans only the width of each burst in time (Table~\ref{tab:pol_properties}). Given the snapshot nature of the correlations, the rather sparse arrays in each run, and the fact that the bursts only covered a fraction of the observed bandwidth, the images from the individual bursts result in elongated fringe patterns, hindering an individual, unambiguous localisation of each burst at the level of the synthesised beam size (see Figure~\ref{fig:localisation-maps}a-d). Therefore, we created a data set that is the combination of the visibilities from all bursts except B1 --- as it was too faint to produce a useful image, and we therefore exclude it from the localization analysis. These data allowed us to unambiguously pinpoint the position of \frb in the field (see Figure~\ref{fig:localisation-maps}e,f). The derived coordinates of \frb in the International Celestial Reference Frame (ICRF)\citep{charlot_2020_aanda} are RA (J2000) $= 9^{\rm h}57^{\rm m}54.69935^{\rm s} \pm 1.2\,\mathrm{mas}$\ Dec (J2000) $= 68^\circ49^\prime0.8529^{\prime\prime} \pm 1.3\,\mathrm{mas}$ (see Methods).
These coordinates coincide with the location of the globular cluster \gc \citep[][]{perelmuter_1995_aj_110}, which is part of the M81 globular cluster system \citep[][]{perelmuter_1995_aj_109}.

\begin{figure*}
        \includegraphics[width=\textwidth]{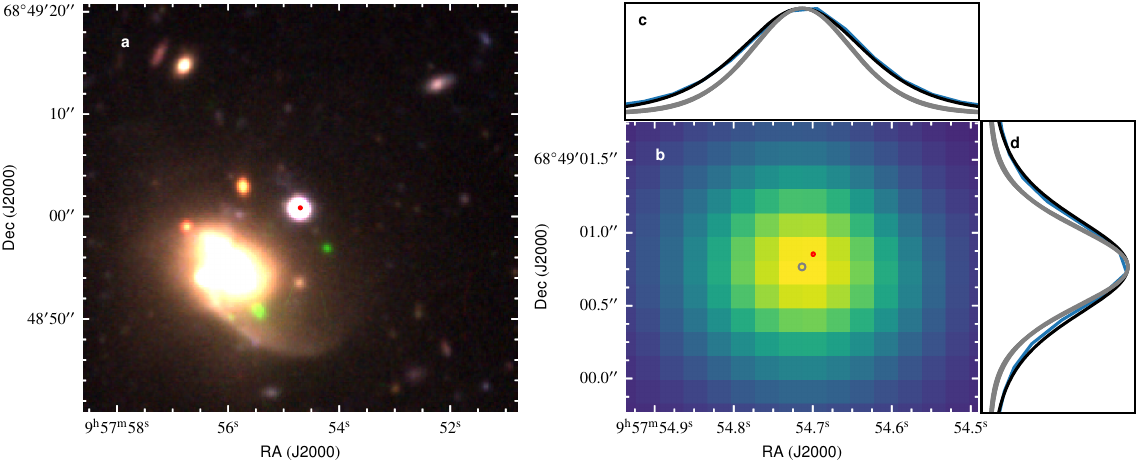}
  \caption{\label{fig:subaru-image} {\bf Optical images of the \frb host and surrounding field.} Left ({\bf a}): 40\arcsec $\times$ 40\arcsec $g^\prime$, $r^\prime$, and $i^\prime$ band image of \gc\ acquired with Hyper Suprime-cam. The small red ellipse is centred at the location of \frb. In panel {\bf b} we show the zoomed-in $r^\prime$ band image of \gc. The grey circle represents the estimated position of the centre of \gc and its 10$\sigma$ uncertainty (dominated by the optical-to-radio reference frame tying). The small red ellipse is the same as in panel \textbf{a}, and also represents the 10$\sigma$ positional uncertainty region of \frb. Panels \textbf{c} and \textbf{d} show cross-sections of the brightness distribution of the cluster (blue solid lines) with the Moffat profile that we fit overlaid in black. Indicated in solid grey lines are the PSFs as measured from stars in the images. Note that scatter in the PSFs is smaller than the linewidth.}
\end{figure*}

Figure~\ref{fig:subaru-image}a shows the position of \frb with respect to \gc in a combined three-colour Subaru image made with $i^\prime$, $r^\prime$, and $g^\prime$ filters mapped to red, green, and blue channels, respectively. The galaxy at the bottom-left of \gc is a background SDSS galaxy at redshift $z=0.194$, i.e. at a $z$ much larger than the maximum possible\citep{bhardwaj_2021_apjl} value ($z<0.03$) based on \frb's dispersion measure. We performed radial fits to the brightness distribution of \gc using a Moffat profile \citep[][]{moffat_1969_aa} in both RA and Dec for all three bands after fitting and subtracting a bilinearly varying background (to account for the presence of the background galaxy). The average position of the centroid of \gc is RA (J2000) $= 9^{\rm h}57^{\rm m}54.7135^{\rm s}~\pm~7$~mas, Dec (J2000) $= 68^\circ49^\prime0.766^{\prime\prime}~\pm~4$~mas (statistical), well in agreement with previous measurements\citep{perelmuter_1995_aj_110} and the position of the source in the {\em Gaia} Early Data Release 3 Catalogue\citep{gaia_2016_aanda,gaia_2021_aanda} (Methods), in which positions are well aligned with the ICRF\citep{gaia_2021_aanda}.

The centre of the FRB localisation is $\approx 116$\,mas offset from the optical centre of light of \gc (Figure~\ref{fig:subaru-image}b, corresponding to $\approx 54\%$ of its effective radius, Methods). 
Given the astrometric uncertainty of the FRB localisation ($\approx 1.3$\,mas) and the optical image registration error with respect to the ICRF ($\approx 15\,\mathrm{mas}$; see Methods), we conclude that \frb is located significantly ($>7\sigma$ confidence level) offset from the optical centre of light of \gc. The optical angular size of \gc (0\farcs77; Methods) in combination with \frb's offset from M81 (19.6\arcmin) and the number of GCs predicted\citep{harris2013} to be part of the galaxy ($300\pm100$), allow us to estimate the probability of chance alignment $P_{\mathrm{cc}}<1.7\times10^{-4}$ (Methods). From such a very low $P_{\rm cc}$ value, we conclude that the association of \frb and \gc is robust.

\begin{figure*}
    \begin{center}
        \includegraphics[width=0.95\textwidth]{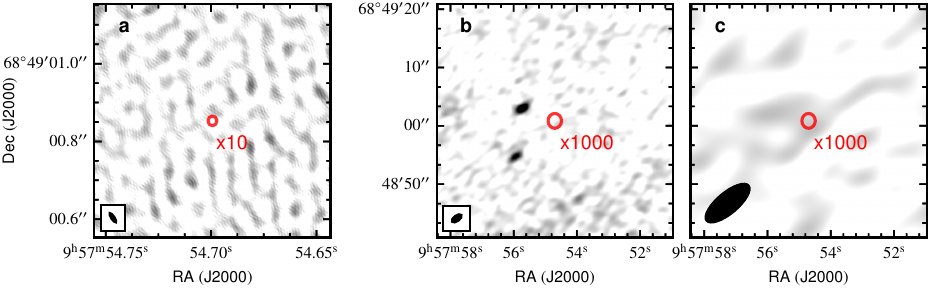}
  \caption{\label{fig:continuum-images}{\bf Continuum maps of the field around \frb.} {\bf a:} 1.4-GHz EVN continuum image after combining the three epochs (EK048B, EK048C, and EK048F); {\bf b:} 1.5-GHz \textit{Realfast}, and  {\bf c:} 340-MHz VLITE continuum image. The red circles indicate the $10\sigma$ (for EVN) and $1000\sigma$ (for \textit{Realfast}, VLITE) positional uncertainty region of \frb. Note the very different scales between the three panels. We clip all values below zero and above $\mathrm{60\,\upmu Jy\,beam^{-1}}$ (EVN), $\mathrm{50\,\upmu Jy\,beam^{-1}}$ (\textit{Realfast}) and $\mathrm{3\,mJy\,beam^{-1}}$ (VLITE) for visualisation purposes. The black ellipse in the bottom left corner of each panel indicates the synthesised beam size and position angle.}
  \end{center}
\end{figure*}

Figure~\ref{fig:continuum-images}a shows a deep continuum map that was created from the combination of the data of all VLBI observations. We find no persistent source at the location of \frb above a $5\sigma$ confidence level (rms background noise level of $\mathrm{10\,\upmu Jy\,beam^{-1}}$). Also shown in Figure~\ref{fig:continuum-images} are continuum images obtained with the Karl G. Jansky Very Large Array (VLA) at $1.5\,$GHz and at $340\,$MHz between December 2020 to January 2021 (Methods). These maps have noise levels of $6.5\,\mathrm{\upmu Jy\,beam^{-1}}$ and $320\,\mathrm{\upmu Jy\,beam^{-1}}$, respectively. Also here, no persistent source is detected at the position of \frb in either of the images. 
For a 1.5\,GHz radio flux density limit of $20\,\mathrm{\upmu Jy}$ ($3\sigma$) and a distance of 3.6\,Mpc, we limit the radio luminosity $L_\nu < 3.1 \times 10^{23}\,\mathrm{erg\,s^{-1}\,Hz^{-1}}$. This luminosity limit is $\sim10^3$ times lower than that of any other extragalactic FRB \citep[][]{marcote_2020_natur} and almost $10^6$ times lower than the radio luminosity of the persistent source in the vicinity of \rone \citep{marcote_2017_apjl}.

We find no evidence of an X-ray source at the location of \frb in archival {\em Chandra} observations (Methods). This results in a  0.5--10\,keV luminosity upper limit of $2\times10^{37}$\,erg\,s$^{-1}$ (3$\sigma$) at the distance of 3.6\,Mpc. A detailed analysis of ongoing X-ray follow-up observations of the region will be presented in Pearlman et al. (in prep.). Similarly, no sources are reported at the location of M81 in any of the {\em Fermi}-LAT catalogues\citep{ajello_2020_apj}. The nearest catalogued source (at a separation of $52^\prime$) is 4FGL\,J0955.7+6940, known to be associated with M82\citep{abdo_2010_apjl}.

Within the context of FRB models that invoke a young, highly magnetized neutron star (NS) powered primarily by the decay of its magnetic field\citep{margalit_2018_apjl, chime_2020_natur_587}, i.e. a magnetar, it is hard to reconcile \frb's association with an old globular cluster using the standard core-collapse supernova formation channel of magnetars.  Instead, because of their extreme stellar densities, globular clusters are known to form short-orbital-period binaries at a high specific rate \citep{hut_1992_pasp,pooley_2003_apjl,verbunt_2003_aspc}.  We thus propose that \frb\ is a magnetar formed via accretion-induced collapse (AIC)\citep{wang_2020_raa} of a white dwarf (WD) or via merger-induced collapse (MIC) of a WD-WD, NS-WD or NS-NS binary\citep{giacomazzo_2013_apjl,schwab_2016_mnras,zhong_2020_apj} --- systems that are common in globular clusters and, like \frb, are found concentrated towards their core\citep{prager_2017_apj} (Methods). The lack of a persistent radio or X-ray source at the position of \frb is expected in an AIC/MIC scenario, as any emission generated during collapse fades on short time scales ($<1$\,yr)\citep{margalit_2019_apj}. 

The globular cluster host of \frb also suggests some alternatives to the magnetar class of FRB models.  \frb could be a compact binary system --- such as a tight WD-NS system in a pre-merger phase or a magnetised NS with a planetary companion \citep{mottez_2014_aa,mottez_2020_aa} --- in which the bodies are interacting magnetically. Similarly, a binary millisecond pulsar with a strong magnetic field formed via AIC and that was subsequently spun-up via accretion \citep{ablimit_2015_apj, ye_2019_apj} could act as an FRB engine. Such a system could also be observable as a low mass X-ray binary (LMXB)\citep{heinke_2003_apj}, as would an accreting black hole (BH). In such an LMXB model, the radio bursts could be generated via magnetic reconnection in a relativistic jet or where the jet shocks with the surrounding medium and creates a synchrotron maser \citep{sridhar_2021_arxiv}. Except for the most luminous LMXBs ($\mathrm{L_X\approx10^{38}\,erg\,s^{-1}}$), our observations cannot rule out such systems. However, none of the $\sim200$ Galactic LMXBs has been seen to generate FRBs.
In some cases, ultra-luminous X-ray sources (ULXs)\citep{kaaret_2017_araa} have been shown to be NSs accreting at hyper-Eddington rates \citep{bachetti_2014_natur}, though some may be systems with a more massive BH primary \citep{webb_2012_sci}. We note that ULXs have been associated with extragalactic globular clusters \citep{dage_2020_mnras} but such systems are ruled out by our X-ray limit unless their luminosity varies in time by more than two orders of magnitude. Additionally, the association with a globular cluster rules out a high mass X-ray binary origin of \frb and the projected offset of $\approx2$\,pc from the centre of light of \gc excludes the association of \frb with, e.g., a massive ($\mathrm{\gtrsim10M_\odot}$) stellar mass BH or a putative intermediate mass black hole at the core of \gc. 

The association of \frb with a globular cluster adds to the diversity of environments in which repeating FRBs  have been found. While \rone resides in a dwarf galaxy\citep{chatterjee_2017_natur}, the host of \rthree is a spiral galaxy\citep{marcote_2020_natur} and \rsixseven was localised to a massive star forming galaxy\citep{ravi_2021_arxiv}. Previously localised repeaters have been associated with nearby star forming regions\citep{bassa_2017_apjl, tendulkar_2021_apjl, piro_2021_arxiv, fong_2021_arxiv}, favouring the core-collapse supernova channel for the formation of a young magnetars, as the rate of AIC and MIC is much lower. The lack of a persistent radio source for all but \rone may suggest a range in the possible ages of such magnetars. In a globular cluster environment, however, the recent core-collapse of a massive star is very unlikely.  Thus, this suggests a diversity in formation channels for magnetars as FRB engines.

\begin{table*}
\caption{\label{tab:pol_properties}Burst properties.}
\begin{tabular}{lcccccc}
\hline
\hline
               &          &  Fluence$\mathrm{^{b,c}}$ & & Peak Flux & Width$\mathrm{^{d}}$ & Gate width $^{\rm e}$ \\
		{Burst}& {MJD $\mathrm{^{a}}$} & { [Jy ms]} & {Peak S/N$\mathrm{^{c}}$} & Density$\mathrm{^{b,c}}$ [Jy] & [$\upmu$s] & { [$\mathrm{\upmu s}$]}\\
		\hline
		B1 &  59265.88304437179  & $0.13\pm0.03$  & 7.8 & $0.9\pm0.2$  &  156 $\pm$ 1 & 290   \\
        B2 &  59265.88600912486  & $0.63\pm0.12$  & 54.9 & $6.6\pm1.3$ &  62\,$\pm$\,1, 93 $\pm$ 0.5 $\mathrm{^{f}}$ & 150\\
        B3 & 59280.69618745651  & $0.52\pm0.10$  & 64.5 & $7.8\pm1.6$ &  46.7 $\pm$ 0.1 &  126\\
        B4 & 59280.80173397988  & $0.71\pm0.14$   & 47.0 & $5.7\pm1.2$ &  117\,$\pm$\,1 & 386\\
        B5 & 59332.50446581106 & 0.09 $\pm$ 0.02 & 11.6 & 1.4\,$\pm$\,0.3 & 56.6 $\pm$ 0.1 & 173\\
		\hline
		\multicolumn{7}{l}{$\mathrm{^{a}}$ Corrected to the Solar System Barycentre and to infinite frequency assuming a dispersion measure of $87.75$ pc\,cm$^{-3}$,}\\
		\multicolumn{7}{l}{ \hspace{0.3cm} reference frequency 1502\,MHz and dispersion constant of 1/(2.41$\times 10^{-4}$)\,MHz$^2$\,pc$^{-1}$\,cm$^{3}$\,s.}\\
		\multicolumn{7}{l}{ \hspace{0.3cm} The times quoted are dynamical times (TDB).}\\
		\multicolumn{7}{l}{$\mathrm{^{b}}$ The receiver temperature of Effelsberg is 20\,K and the telescope gain is 1.54 K\,Jy$^{-1}$. We additionally consider }\\
		\multicolumn{7}{l}{ \hspace{0.3cm}a sky background temperature of 0.8\,K, by extrapolating from the 408\,MHz map \citep{remazeilles_2015_mnras},}\\
		\multicolumn{7}{l}{\hspace{0.3cm}using a spectral index of $-$2.7 \citep{reich_1988_aa}, and 3\,K from the cosmic microwave background}\\
		\multicolumn{7}{l}{\hspace{0.3cm}\citep{mather_1994_apj}. We take a conservative 20\,\% error on these measurements, arising due to the uncertainty}\\
		\multicolumn{7}{l}{\hspace{0.3cm}in the Effelsberg receiver temperature and gain.}\\
		\multicolumn{7}{l}{$\mathrm{^{c}}$ Computed using the frequency range over which the burst is bright.}\\
		\multicolumn{7}{l}{$\mathrm{^{d}}$Defined as $1/\sqrt{2}$ multiplied by the full-width at half-maximum of the autocorrelation function (ACF).} \\
		\multicolumn{7}{l}{ \hspace{0.3cm}  Note we use the $\pm2\sigma$ burst width region to determine the burst fluence (see Nimmo et al. (submitted) for details).} \\
		\multicolumn{7}{l}{$\mathrm{^{e}}$ Width of the gate used for the interferometric correlation of each burst. For B2 the gate was centered on the first, brighter} \\
		\multicolumn{7}{l}{\hspace{0.3cm}component to maximise S/N.} \\
		\multicolumn{7}{l}{$\mathrm{^f}$ Width per burst component.} \\
\end{tabular}
\end{table*}

\clearpage
\clearpage 

\section*{Methods} 

\subsection*{Observations and Data reduction}
\label{sec:observations}

\subsubsection*{VLBI observations} 

As part of our ongoing very long baseline interferometry (VLBI) campaign called PRECISE (Pinpointing REpeating ChIme Sources with the EVN), we observed the field centred on the coordinates from the discovery\citep{bhardwaj_2021_apjl} of \frb: RA (J2000) $= 09^{\rm h}57^{\rm m}56.688^{\rm s}$,\ Dec (J2000) $= 68^{\circ}49^{\prime}31.8^{\prime\prime}$. The quoted 90\% confidence interval\citep{bhardwaj_2021_apjl} of radius $\sim1.5$\arcmin\ is well covered by the field of view of the \emph{ad-hoc} interferometric array we used to observe, which is comprised of dishes that are also part of the EVN. We observed the field 15 times between February and May 2021 (Table~\ref{tab:all-precise-runs-on-m81}), with burst detections in the data that were taken on 2021 February 20 UT 17:00--22:00 (project code PR141A), 2021 March 7 UT 15:45--20:45 (PR143A), and 2021 April 28 UT 11:00--22:00 (PR158A). We observed in the 21-cm band ($\sim1.4$\,GHz) with slightly different array setups in each run. Depending on the capabilities at each station, we recorded either 128 or 256\,MHz of total bandwidth, divided into 8 or 16 subbands of 16\,MHz each. The participating stations and their respective frequency coverage are summarised in Table~\ref{tab:stations}. We ran regular phase-referencing observations with a cycle time of 7.5\,min; that is, 5.5\,minutes on target and 2.0\,minutes on the phase calibrator (J0955$+$6903, at $\approx 0.32^{\circ}$ separation). In total, we spent 2.93\,hrs on the target field in each of PR141A and PR143A, and 6.73\,hrs in PR158A.
The calibrator source J1048$+$7143 served as fringe finder and bandpass calibrator and was observed twice for 5\,minutes in each run. For verification and single-dish time-domain calibration purposes we observed the pulsar B0355$+$54 for 2\,minutes in PR141A and PR143A. In PR158A the pulsar B2255$+$58 was observed for 5\,minutes for the same reason.

We recorded raw voltages (`baseband' data, dual circular polarisation, 2-bit sampling) in VDIF \citep[][]{whitney_2010_ivs} or Mark5B\citep[][]{whitney_2004_evn} format at all stations. At Effelsberg we also recorded total intensity filterbanks with the PSRIX pulsar backend \citep[][]{lazarus_2016_mnras} in parallel to the voltage data. These data span the frequency range of $1255$--$1505$\,MHz with a time and frequency resolution of 102.4\,\us and 0.49\,MHz, respectively. Similarly, at SRT we also recorded in parallel baseband data and total intensity filterbanks (with the local Digital Filter Bank system, DFB). These DFB-filterbanks have a time and frequency resolution of 128\,\us and 1\,MHz, respectively, covering the frequency range of 1140.5--2163.5\, MHz out of which we search the usable range of 1210.5--1739.5\,MHz.

The data from both Effelsberg and SRT were searched for bursts in two independent pipelines. The baseband data were processed with the pipeline outlined in \citet{kirsten_2021_natas}, which converts the raw voltages to filterbank format (in this case, total intensity with time resolution 64\,\us\ and frequency resolution 125\,kHz) and searches those with {\tt Heimdall} within $\pm 50\,$\dmunit of the expected $\mathrm{ DM=87.818\,}$\dmunit, as found by \citet{bhardwaj_2021_apjl}. The burst candidates are classified as either radio frequency interference (RFI) or potential FRBs using the neural network classifier FETCH \citep[][]{agarwal_2020_mnras}. The filterbanks as recorded with the respective pulsar backends were searched with a pipeline that uses the PRESTO suite of tools \citep[][]{ransom_2001_phdt, ransom_2011_ascl} and a classifier based on work by \citet{michilli_2018_mnras}.

Correlation of the data was performed at the Joint Institute for VLBI ERIC (JIVE; in the Netherlands) under proposal EK048 (with codes EK048B for PR141A, EK048C for PR143A, and EK048F for PR158A). In total, we ran three correlator passes using the software correlator SFXC \citep[][]{keimpema_2015_exa}: in the first run we correlate all scans containing the calibrator sources with a standard 2-s time integration and 64 channels per 16-MHz subband. A second correlation pass was performed only on the data containing bursts (their arrival times were determined via the search described above), with a higher frequency resolution (8\,192 channels per subband) and time resolution (the correlation gates around the bursts were chosen manually to optimise the signal-to-noise for each burst, Table~\ref{tab:pol_properties}).
The strongest bursts allowed a direct fringe fit on their data, which provided a first estimate of their sky position to a level of a few arcseconds: RA (J2000) $\approx 09^{\rm h}57^{\rm m}54.8^{\rm s}$,\ Dec (J2000) $\approx 68^\circ49^\prime03^{\prime\prime}$.
We then used this position to re-correlate the burst data at the same frequency resolution as in the first correlation pass. This allowed us to directly apply the calibration performed on the first pass (containing the calibrator sources) to the data containing bursts.

The full data reduction was performed using the Astronomical Image Processing System \citep[{\tt AIPS}, ][]{greisen_2003_assl} and {\sc Difmap} \citep{shepherd_1994_baas} following standard procedures. \emph{A-priori} amplitude calibration was performed using the known gain curves and system temperature measurements recorded at each station during the observation.
These data were not available for Toru\'{n} and Irbene in EK048B, and for Urumqi in EK048C. We used nominal system equivalent flux density values and flat gain curves to perform the amplitude calibration in these cases. In EK048C, the data from Urumqi were flagged due to the impossibility to properly convert the locally-recorded linear polarizations into circular ones (as in the rest of the antennas).
Ionospheric dispersive delays were calculated from maps of total electron content provided by the global positioning system satellites and removed via the {\tt TECOR} task in {\tt AIPS}.
We first corrected for the instrumental delays and bandpass calibration using the calibrator source J1048+7143, and thereafter fringe-fit the data using all calibrator sources. The phase calibrator (J0955+6903; displaying a peak brightness of $\sim 73\,\mathrm{mJy\,beam^{-1}}$) was then imaged and self-calibrated to improve the final calibration of the data (for both phases and amplitudes). A common source model for the phase calibrator was used to improve the calibration of both epochs.  The obtained solutions were then transferred to the burst data, which were finally imaged.

In the third, and final, correlator pass we processed the entire observation, correlating all scans from the target with the same time and frequency setup as in the first correlator pass, in order to obtain a deep continuum image of the field.

\begin{table*}[h!t]
\begin{center}
\caption{\label{tab:all-precise-runs-on-m81}Time ranges of PRECISE runs targeting \frb between February and May 2021}
\begin{tabular}{lccc}
\hline
Observation  & EVN & & \\
Project code & Project code & Start MJD & Stop MJD \\
\hline
\hline
PR141A$^\mathrm{a}$ & EK048B & 59265.708 & 59265.916 \\
PR143A$^\mathrm{a}$ & EK048C & 59280.656 & 59280.864 \\
PR144A & & 59283.792 & 59284.000 \\
PR145A & & 59289.750 & 59289.958 \\
PR146A & & 59295.667 & 59295.875 \\
PR153A & & 59314.887 & 59314.972 \\
PR158A$^\mathrm{a}$ & EK048F & 59332.458 & 59332.916 \\
PR159A & & 59336.708 & 59337.000  \\
PR160A & & 59341.833 & 59342.072 \\
PR161A & & 59344.771 & 59344.875 \\
PR162A & & 59346.646 & 59346.895 \\
PR163A & & 59347.417 & 59347.625 \\
PR164A & & 59351.917 & 59352.166 \\
PR165A & & 59358.917 & 59359.166 \\
PR166A & & 59360.708 & 59360.916 \\

\hline
\multicolumn{3}{l}{$^{\mathrm{a}}$Epoch with detection}\\
\end{tabular}
\end{center}
\end{table*}

\begin{table*}[h!t]
\begin{center}
\caption{\label{tab:stations}Setups at the different stations during observations used in the analysis}
\begin{tabular}{lccc}
\hline
Telescope & Frequency coverage {[}MHz{]} & Station project code & EVN project code \\
\hline
\hline
Effelsberg (Ef) & $1254-1510$ & 94-20 & EK048B/C/F$^{\mathrm{b}}$ \\
Medicina (Mc) & $1350-1478$ & 44-20 & EK048B/C/F$^{\mathrm{c}}$ \\
Noto (Nt) & $1318-1574$ & 44-20 & EK048\phantom{B/}C\phantom{/F}\phantom{$^{\mathrm{b}}$}\\
Irbene (Ir) & $1382-1510$ & -- & EK048B/C/F\phantom{$^{\mathrm{b}}$} \\
Toru\'{n} (Tr) & $1254-1510$ & DDT$^{\mathrm{a}}$ & EK048B/C/F$^{\mathrm{b}}$ \\
Westerbork (Wb) & $1382-1510$ & DDT$^{\mathrm{a}}$ & EK048B/C/F\phantom{$^{\mathrm{b}}$} \\
Urumqi (Ur) & $1382-1510$ & DDT$^{\mathrm{a}}$ & EK048\phantom{B/}C/F\phantom{$^{\mathrm{b}}$} \\
Sardinia (Sr) & $1360-1488$ & 44-20 & EK048\phantom{B/C/}F\phantom{$^{\mathrm{b}}$} \\
Onsala (O8) & $1382-1510$ & DDT$^{\mathrm{a}}$ & EK048B\phantom{/C/}F$^{\mathrm{b}}$ \\
Badary (Bd) & $1382-1510$ & DDT$^{\mathrm{a}}$ & EK048\phantom{B/C/}F\phantom{$^{\mathrm{b}}$} \\
Svetloe (Sv) & $1382-1510$ & DDT$^{\mathrm{a}}$ & EK048\phantom{B/C/}F\phantom{$^{\mathrm{b}}$} \\
Zelenchukskaya (Zc) & $1382-1510$ & DDT$^{\mathrm{a}}$ & EK048\phantom{B/C/}F\phantom{$^{\mathrm{b}}$} \\
VLA-VLITE   & $320-384$ & 20B-280 & --\\
VLA-\textit{Realfast} & $1300-1700$ & 20B-280 & --\\
\hline
\multicolumn{4}{l}{$^{\mathrm{a}}$Director's Discretionary Time}\\
\multicolumn{4}{l}{$^{\mathrm{b}}$Only stations recording the part of the band where the burst (B5) was detected with significant emission.}\\
\multicolumn{4}{l}{$^{\mathrm{c}}$Only one subband overlapping with the part of the band where the burst (B5) was detected with significant emission.}\\
\end{tabular}
\end{center}
\end{table*}

\subsubsection*{VLA} 

We observed the field of \frb with the Karl G. Jansky Very Large Array (VLA) as part of a program to localise repeating FRBs identified by CHIME/FRB (program 20B-280). The field was observed in five 1-hour blocks from 2020 December 29 through 2021 January 18. Data were recorded in parallel at both $1.5\,$GHz and $340\,$MHz. The VLA antennas were arranged in the A configuration, which provides baseline lengths up to 30\,km and a typical spatial resolution of 1.3\arcsec\ and 6.0\arcsec\ at $1.5\,$GHz and $340\,$MHz, respectively. The total on-target integration time amounts to 200\,minutes.

\paragraph{Realfast} 

Visibility data from the VLA observations were recorded with a sampling time of 3 seconds from 1 to 2\,GHz. Simultaneously, a copy of the data with sampling time of 10\,ms was streamed into the \textit{Realfast} system \citep{law_2018_apjs}. We used \textit{Realfast} to search this data stream for FRBs in real time. The typical $1\sigma$ sensitivity of the VLA is 5\,mJy\,beam$^{-1}$ in 10\,ms.

The standard (slow) visibility data were analyzed to search for persistent emission associated with the FRB location. We calibrated the data with the VLA calibration pipeline (version 2020.1) using 3C147 as a flux calibrator. Calibrated visibilities for all five epochs were combined and imaged with CASA (version 6.1).

We imaged the data with robust weighting of 0.5 and removed baselines shorter than $100\lambda$ to reduce the effects of RFI. This produced an image with a synthesized beam size of 2\arcsec\ by 1\arcsec\ with position angle of $126^{\circ}$. The sensitivity in the combined image is $\mathrm{6.5\,\upmu Jy\,beam^{-1}}$, which is consistent with expectations given the usable bandwidth of $400\,$MHz.

\paragraph{VLITE} 

The VLA Low-band Ionosphere and Transient Experiment \citep[VLITE; ][]{polisensky_2016_apj, clarke_2016_spie} is a commensal instrument on the VLA that records and correlates data from a 64-MHz subband at a central frequency of $\sim340\,$MHz.  It operates on up to 18 antennas during nearly all regular VLA operations. All VLITE data were processed within the VLITE-Fast GPU-based real-time system to search the incoming voltage stream for dispersed transients \citep[][]{bethapudi_2021_rnaas}. 
For imaging purposes, primary calibration and editing for each day of visibility data were carried out with the automated VLITE processing pipeline.  Due to radio interference from satellites at the upper end of the band, the final usable bandwidth was 38.2\,MHz centred at 340.85\,MHz. The data were subsequently combined, imaged and self-calibrated in amplitude and phase using the {\tt Obit} task {\tt MFImage} \citep{cotton_2008_pasp}.  In order to reduce artefacts from the bright extended radio galaxy M82 located $\sim1^{\circ}$ northwest of the target position, baselines shorter than 4.0\,k$\lambda$ were removed at this point. The final image was created in {\tt WSClean} \citep{offringa_2014_mnras}, and corrected for the offset primary beam response of VLITE.  The image has an rms of $\mathrm{320\,\upmu Jy\,beam^{-1}}$, and a beam of 10.1\arcsec\ by 3.6\arcsec\ at a position angle of $132^{\circ}$.

\subsubsection*{Archival optical and high energy data}

\paragraph{Hyper Suprime-Cam} 
The field around M81 has been well-observed over the years by multiple telescopes. We retrieved archival data from the Hyper Suprime-Cam on the 8.1-m Subaru telescope \citep{miyazaki_2012_spie} using the SMOKA interface.  We chose images with seeing better than 0\farcs7. We processed the $g^\prime$, $r^\prime$, and $i^\prime$ band images with the \texttt{hscpipev8.4} pipeline \citep{bosch_pasj_2018}. The pipeline uses the PanSTARRs catalogue \citep[PS1; ][]{flewelling_2020_apjs} as an astrometric and photometric reference. The typical astrometric residuals were $\sim 50$--60\,mas, which is the uncertainty we have assumed to tie the optical and the radio reference frames.

\paragraph{{\em Gaia}} 
\gc also appears in the {\em Gaia} Early Data Release 3 Catalogue\citep{gaia_2016_aanda,gaia_2021_aanda} with Source ID 1070264274879949184, and position RA (J2000) $= 9^{\rm h}57^{\rm m}54.71402^{\rm s} \pm 1.6\,\mathrm{mas}$,\ Dec (J2000) $= 68^\circ49^\prime0.7775^{\prime\prime} \pm 1.7\,\mathrm{mas}$. This position is  consistent (within $<3\sigma$ confidence level) with the one we have derived from the Hyper Suprime-Cam data. The observed offset allowed us to estimate the possible systematic uncertainties in the optical image registration error ($15\,\mathrm{mas}$; by adding in quadrature the observed offset between the the {\em Gaia} position and the one we determined, plus the uncertainties on the positions).

\paragraph{{\em Chandra} X-ray Observatory} 

Several deep archival X-ray observations are available for the field around M81 from {\em XMM} and {\em Chandra}. We selected the archival observation with the longest exposure time that covers the location of \frb, a 26\,ks {\em Chandra} observation \citep[Obs. ID 9540, taken with ACIS in FAINT mode, ][]{garmire_2003_spie} to probe for an X-ray source. The data were reduced using CIAO version 4.12 \citep[][]{fruscione_2006_spie} following standard procedures. As the source was located about 14\arcmin\ off-axis, events were extracted in a large 10\arcsec-radius region centred on the position of \frb. We also extract events from a 60\arcsec-radius region away from the source at a similar off-axis angle to estimate the background count rate.
The X-ray count-rate in the source extraction region, $4.4\times10^{-6}$ counts\,s$^{-1}$\,arcsec$^{-2}$, is consistent with the background region rate of $3.9\times10^{-6}$ counts\,s$^{-1}$\,arcsec$^{-2}$. To place a limit on an X-ray source at the location of \frb\ we use the Bayesian method of \citet{kraft_1991_apj}, which results in a 0.5--10\,keV source count rate upper limit of $1\times10^{-3}$ counts\,s$^{-1}$ (3$\sigma$).  Taking into account the spectral response for the off-axis location of the source (via an ancillary response file created by the CIAO tool {\tt specextract}), and assuming a photoelectrically absorbed power-law source spectrum with a spectral index of $\Gamma=2$ and a hydrogen column density of $N_\mathrm{H} = 10^{21}$\,cm$^{-2}$, this count rate limit corresponds to a 0.5--10\,keV absorbed flux upper limit of $1\times10^{-14}$\,erg\,cm$^{-2}$\,s$^{-1}$.

\paragraph{{\em Fermi}-LAT} 

The Large Area Telescope (LAT) onboard the \textit{Fermi} satellite provides a uniform sensitivity survey of the whole sky in the energy range between 100~MeV and 300~GeV.  We searched all of the publicly available catalogues for counterparts up to the latest published\citep{ballet_2020_arxiv} release, 4FGL-DR2, with null results. However, even the most luminous known Galactic globular cluster (Terzan5), whose luminosity is $(42.4 \pm 1.5) \times 10^{34}$ erg s$^{-1}$ in the $0.1$--$100$\,GeV energy range\citep{abdo_2010_AandA}, would have a gamma-ray flux of only  $(2.00 \pm 0.07) \times 10^{-16}$\,erg\,cm$^{-2}$\,s$^{-1}$ at the distance of M81.  This is nearly three orders of magnitude dimmer than the faintest source detected in the 4FGL-DR2 catalogue.

\subsection*{Analysis}

\subsubsection*{Dispersion measure refinement} 

To refine the DM for further analysis, we maximised the signal to noise ratio (S/N) of a very narrow spike in B3\citep{nimmo_2021_arxiv} (Figure~\ref{fig:bursts}g) and find $\mathrm{DM=87.7527\pm0.0003\,}$\dmunit. This value is comparable to, but formally deviates from $\mathrm{ DM=87.818\pm0.007\,}$\dmunit\ found by \citet{bhardwaj_2021_apjl}, where a weighted average of three bursts was used. Possible explanations for this discrepancy are the lack of short-timescale structure in the CHIME/FRB bursts\citep{bhardwaj_2021_apjl}, the potential for non-dispersive time-frequency drifting\citep{hessels_2019_apjl} or a time-variable DM.

\subsubsection*{Milliarcsecond localization of \frb} 

We imaged both the individual bursts separately, as well as a data set produced by the combination of all individual burst visibilities.
Figure~\ref{fig:localisation-maps}a-d display the dirty maps (i.e.\ the inverse Fourier Transform of the visibilities without applying any deconvolution, i.e. `CLEANing') of the bursts that were detectable in the correlated data, using a natural weighting of the data. B1 was too faint to produce a useful image, and we therefore exclude it from the localisation analysis. B5 was only detected in the lower half of the observed band, where most of the antennas were not recording (see Table~\ref{tab:stations}). We therefore only used data from this part of the band. Figure~\ref{fig:localisation-maps}e shows the dirty map of the combined data of the visibilities from all bursts. In this map we obtained an emission pattern consistent with the one expected from the dirty beam (the inverse Fourier Transform of a point-like source), allowing us to unambiguously identify the position of \frb. The observed emission reached a $12\sigma$ confidence level, with the secondary sidelobes in the fringe pattern being $\sim 66\%$ of the peak emission. This provided a robust localization in the map, as it would require a noise fluctuation of $\gtrsim 7\sigma$ to produce such peak emission.  We also conducted different approaches during the imaging of the data: different weighting schemes, and selecting different subsets of antennas in the array. The derived position of \frb was robust across all these approaches. Figure~\ref{fig:localisation-maps}f displays the final, `CLEANed', image of the combined bursts.

The final coordinates of \frb are RA (J2000) $= 9^{\rm h}57^{\rm m}54.69935^{\rm s} \pm 1.2\,\mathrm{mas}$,\ Dec (J2000) $= 68^\circ49^\prime0.8529^{\prime\prime} \pm 1.3\,\mathrm{mas}$. We note that the quoted uncertainties reflect the statistical uncertainties from the measured position of \frb ($0.7$ and $0.4$\,mas in RA and DEC, respectively), the uncertainties in the absolute International Celestial Reference Frame position of the phase calibrator (J0955$+$6903; 0.11\,mas), and the systematic uncertainty associated with the phase-referencing technique\citep{kirsten_2015_aanda} of $\sim 0.9$ and $1.2$\,mas, in RA and DEC, respectively.

We combined the continuum VLBI data from the three epochs to produce a deep image of the field around \frb to search for persistent emission. No significant sources above a $5\sigma$ confidence level (with an rms of $10\,\mathrm{\upmu Jy\,beam^{-1}}$) are detected on milliarcsecond scales (Figure~\ref{fig:continuum-images}a).
In the VLA data taken at $1.5\,$GHz (Figure~\ref{fig:continuum-images}b), we did identify two sources with peak brightness of $110\,\mathrm{\upmu Jy\,beam^{-1}}$ and $73\,\mathrm{\upmu Jy\,beam^{-1}}$, and offset by $6\arcsec$ and $9\arcsec$, respectively. The projected density of radio sources of this brightness is roughly 1\,000 to 3\,000 per square degree \citep[][]{condon_2012_apj}. Therefore, we expect between $1$--$3$ sources within $1\arcmin$ of \frb by chance. The closer of the two nearby radio sources is within $0.2\arcsec$ of a PS1 source with $i = 21.3\,\mathrm{mag}$. This host galaxy has a photometric redshift in the PS1 STRM catalogue \citep{beck_2021_mnras} of $0.67\pm0.2$, so it is most likely a background galaxy. The other identified radio source has no PS1 counterpart. We note that neither of these two sources exhibit significant compact emission on milliarcsecond scales.

\subsubsection*{\gc and chance coincidence probability} 
We measured the full-width at half maximum (FWHM; ``seeing'') of the coadded $i^\prime$, $r^\prime$, and $g^\prime$ images of \gc to be 0\farcs63, 0\farcs57, and 0\farcs62, respectively, using profile fits to bright, isolated stars. The brightness distribution of \gc has an FWHM of 0\farcs77, 0\farcs70, and 0\farcs75 in the same three images. Subtracting the FWHMs of the isolated stars in quadrature, we estimate that the intrinsic FWHM of \gc is about 0\farcs42, corresponding to about 7.4\,pc at a distance of 3.63\,Mpc. 

To estimate the probability of chance coincidence\citep{bloom_2002_aj} for an M81 globular cluster in the FRB localisation region, we use a circular localisation region with a radius ($R$) = $\max \{2R_{\rm{eff}}$ of \gc, maximum seeing-limited size of \gc in  the Hyper Suprime-images\}  = 0\farcs77.
Perelmuter \& Racine\citep{perelmuter_1995_aj_109} parameterized the projected areal number density ($\rho_{\rm GC}/ {\rm arcmin}^{2}$) of the M81 globular clusters as a function of their angular offset from M81 ($r$, in arcmin) as, $\log_{10}(\rho_{\rm GC}) = -2.07 \times \log_{10}(r) + 0.82 \pm 0.05$, for $0 \leq \log_{10}(r) \leq 1.4$. At the offset of \frb from M81 ($19.6\arcmin$), $\rho_{\rm GC} = 0.014\;{\rm arcmin}^{-2} \approx 3.8\times 10^{-6}\; {\rm arcsec}^{-2}$. Assuming a Poisson distribution of M81 globular clusters at $r = 19.6\arcmin$, the probability of finding at least one globular cluster by chance within a radius $R$ is given by $P_{\rm cc} = 1 - \exp(-\pi R^{2} \rho_{\rm GC}) \approx 7 \times 10^{-6}$. A more conservative estimate of $P_{\rm cc}$ can be derived by assuming that all the predicted $300 \pm 100$ M81 globular clusters \citep{harris2013} are uniformly distributed within the angular area of radius = $19.6\arcmin$ so that $\rho_{\rm GC} = 4.6$--$9.2 \times 10^{-5}$, and consequently $P_{\rm cc}$ = 0.85$- 1.7 \times 10^{-4}$. From such a very low $P_{\rm cc}$ value, we conclude that the association of \frb and \gc is robust.         

\subsubsection*{Modelling of \gc} 

\begin{figure*}[h]
    \centering
    \includegraphics[width=0.95\textwidth]{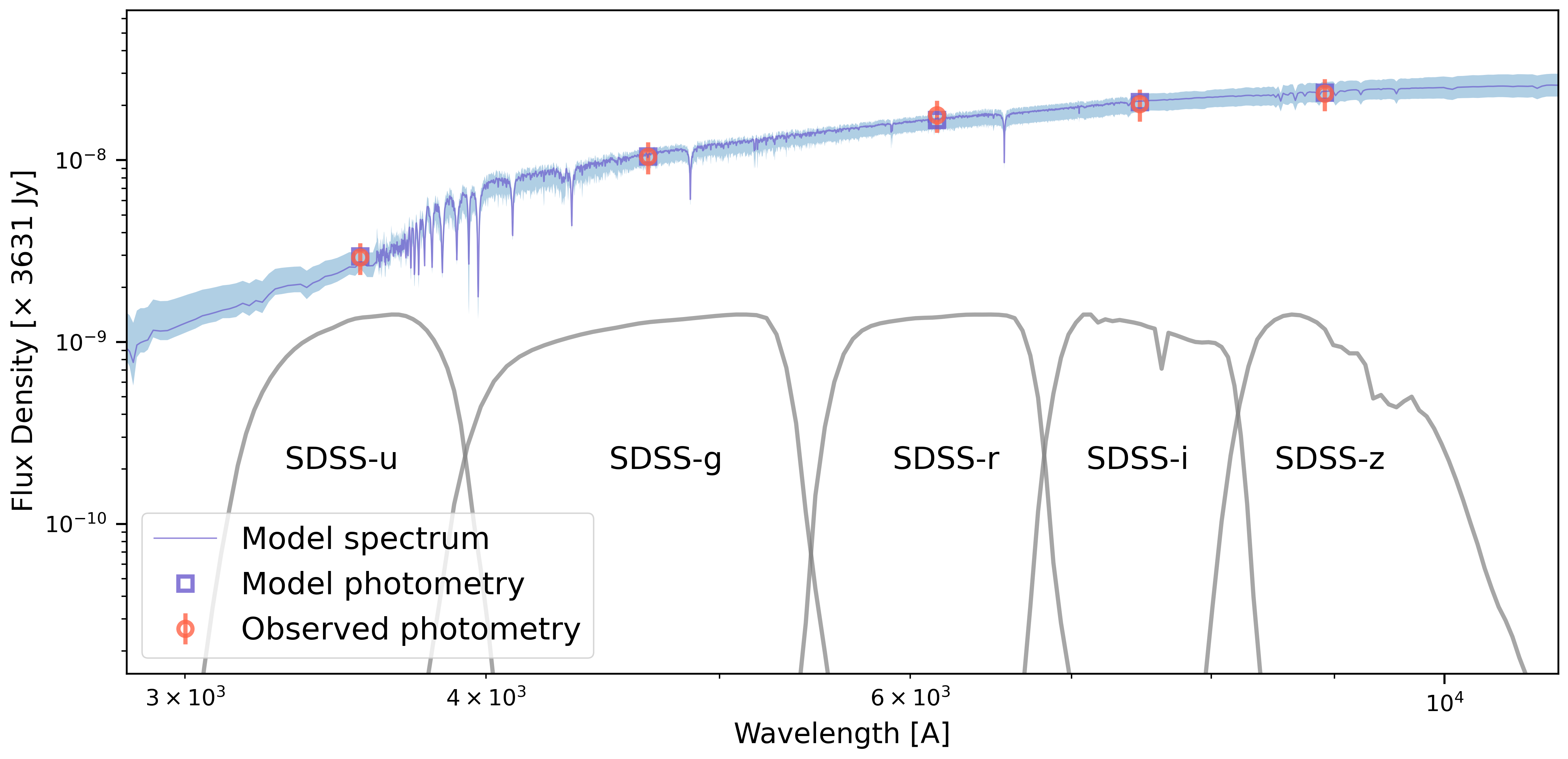}
    \caption{\textbf{Modelling the SED of \gc.} The Milky Way extinction corrected flux densities of \gc in different wavelength bands are plotted along with the best fit \texttt{Prospector} model spectrum. To assess the quality of the \texttt{Prospector} model, the modelled and actual photometric data are also shown. The best fit model profile is used to estimate the physical properties of \gc stated in Table~\ref{tab:gc-properties}. Finally, the shaded region around the best fitted profile is the one-$\sigma$ uncertainty region.}
    \label{fig:sed}
\end{figure*}

\begin{figure*}[h]
    \centering
    \includegraphics[width=0.95\textwidth]{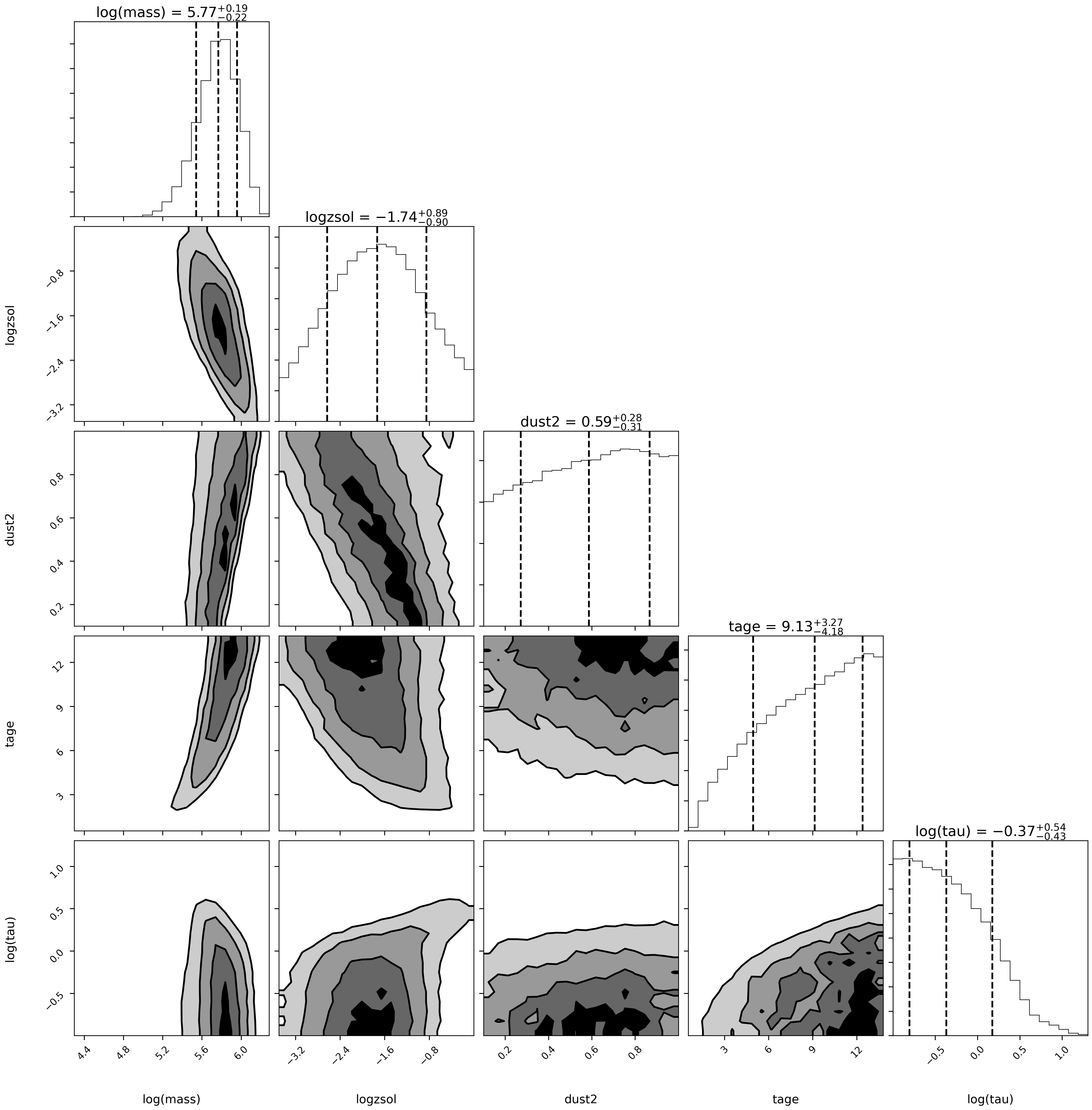}
    \caption{\textbf{MCMC simulation corner plot}, where the posterior probability distributions are shown for each of the five model parameters along the diagonal panels, and the correlations between model parameter posteriors are shown along the columns. Above each probability distribution, the median of the parameter posterior is printed, along with the one-$\sigma$ error bars.}
    \label{fig:mcmc}
\end{figure*}

To estimate important physical properties of \gc, such as its stellar mass, metallicity, stellar population age and V-band extinction, we used the {\tt Prospector} code \citep{Leja2017,prospect2019} for stellar population inference. We modelled the SDSS photometry (Table~\ref{filter:maggies}, from the SDSS DR12 catalogue\citep{aaa2015}) of \gc and fit a five-parameter (Table~\ref{tab:sfhmodel}) `delayed-tau' model for the star formation rate $\mathrm{SFR}(t) \propto t\exp(-t /\tau)$, where $t$ is the time since the formation epoch of the galaxy, and $\tau$ is the characteristic decay time of the star-formation history of \gc \citep{simha2014,carnall2019ApJ}. As we are modelling a globular cluster, we did not include nebular line emission but only enabled a dust emission model\cite{draine2007} in our fitting. The best-fit spectral energy distribution (SED) profile of \gc is shown in Figure~\ref{fig:sed}. {\tt Prospector} also enables Markov Chain Monte Carlo (MCMC) sampling of the posterior to estimate uncertainty in the best-fit values of the physical properties of \gc. We show the corner plot of the MCMC-analysis in Figure~\ref{fig:mcmc} and list the results in Table~\ref{tab:gc-properties}. Using the relation\citep{bertelli1994}, [Fe/H] = 1.024 $\times$ log(Z/Z$_{\odot}$), we estimate the [Fe/H] of \gc = $-1.83^{+0.86}_{-0.87}$, which is in good agreement with earlier estimates\citep{ma_2007_pasp, perelmuter_1995_aj_110}. We can get an estimate of the velocity dispersion ($\sigma_{\rm r}$) of \gc using the virial theorem: $\sigma_{\rm r} \sim \sqrt{\rm{2GM/3R_{eff}}}$, where G is the gravitational constant. Using the M and R$_{\rm eff}$ values of \gc from Table~\ref{tab:gc-properties}, we estimate $\sigma_{\rm r} \sim 22\,\mathrm{km\,s^{-1}}$.

\subsubsection*{Possible MIC-models for the formation of \frb} 
The most likely of the MIC-models is that of a merging WD-WD system as those dominate the cores of globular clusters\citep{kremer_2021_arxiv}, while a NS-NS progenitor system is less probable \citep{ye_2020_apjl}. A typical globular cluster with a total mass of $\mathrm{\sim2\times10^5\,M_\odot}$ can host $10-20$ MSPs formed via AIC or MIC\citep{ye_2019_apj}. At birth, such NSs would be extreme objects with high rotation rates and strong magnetic fields, potentially capable or generating FRBs. We note that the metallicity limit ($\mathrm{log[Fe/H] > - 0.6}$) set for the formation of young NSs in globular clusters\citep{boyles_2011_apj} is much higher than the metallicity that we estimate for \gc\ ($\mathrm{log[Fe/H] = - 1.83}$, Table~\ref{tab:gc-properties}). Nevertheless, non-recycled pulsars have been found towards some globular clusters\citep{hessels_2015_aska, lyne_1996_apjl,boyles_2011_apj}. These young pulsars are formed at a comparable rate to MSPs in globular clusters, but they have much shorter active lifetimes (Myr versus Gyr).

\subsubsection*{Constraints on the Galactic halo DM and RM} 
\begin{figure*}
    \begin{center}
        \includegraphics[width=0.5\textwidth]{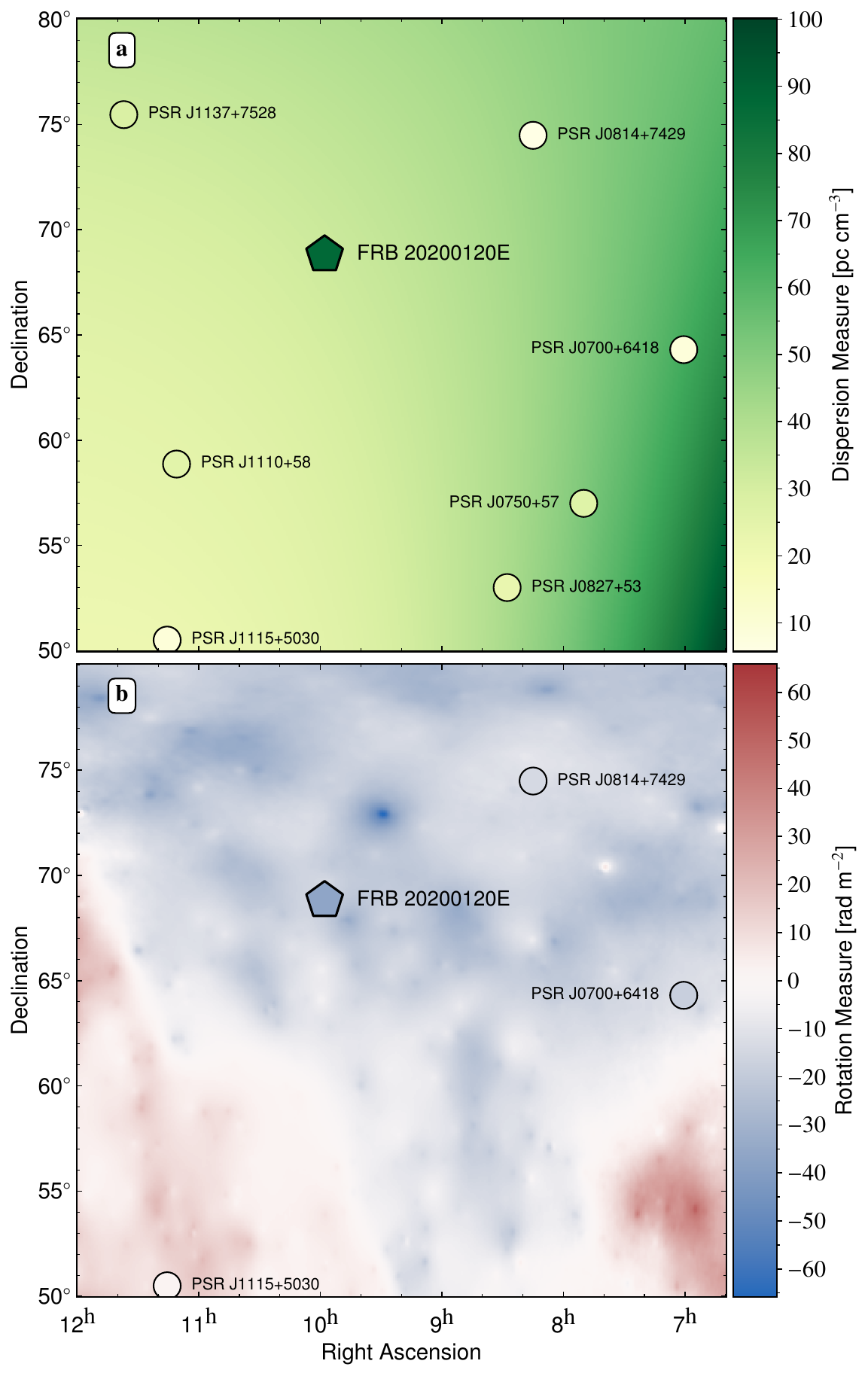} 
  \caption{\label{fig:dm_rm_map} {\bf Dispersion measure (DM) and rotation measure (RM) maps around \frb:} Panel {\bf a} shows the expected Galactic DM contribution (background) according to the YMW16 model \citep[disk contribution only,][]{yao_2017_apj}, the DM of \frb (pentagon) and the DMs of known pulsars from the ATNF Pulsar Catalogue in this field \citep[circles,][]{hobbs_2004_iaus}. Panel {\bf b} shows the physical Galactic Faraday depth $\phi_{\text{g}}$ \citep[background,][]{hutschenreuter_2021_arxiv}, the RM of \frb (pentagon) and galactic pulsars with a known RM (circles). We assume that the RM\citep{bhardwaj_2021_apjl} of \frb is $\mathrm{-36.9\,rad\,m^{-2}}$ (also see Nimmo et al. (submitted) for details).}
  \end{center}
\end{figure*}

Since \frb\ is extragalactic, its dispersive delay is probing the full extent of the Milky Way ISM ($\mathrm{DM_{ISM}^{MW}}$) and halo ($\mathrm{DM_{Halo}^{MW}}$) along this line of sight.  There are also extragalactic contributions from the intergalactic medium $\mathrm{DM_{IGM}}$ and the host galaxy ($\mathrm{DM_{M81}}$), leading to the total observed $\mathrm{DM=DM_{ISM}^{MW}+DM_{Halo}^{MW}+DM_{IGM}+DM_{M81}}$.
The intergalactic medium between the Milky Way and M81 contributes on the order\citep{macquart_2020_natur} of DM$_{\rm IGM} \sim 1$\,\dmunit.
To estimate DM$_{\rm M81}$, we only consider the contribution from the halo of M81, as both the M81 disk and the globular cluster make negligible contributions to the measured DM. The M81 halo DM contribution is estimated\citep{bhardwaj_2021_apjl} to be $\sim 15$--$50$~\dmunit depending on the choice of a halo density profile and baryon fraction in the M81 halo.    
For the Galactic contribution, models of the Milky Way ISM predict DM$^{\rm MW}_{\rm ISM}=30$--$40$\,\dmunit\ in this direction \citep{cordes_2002_arxiv,yao_2017_apj} (with a shallow spacial gradient in the region, Figure~\ref{fig:dm_rm_map}). Using these estimates, we limit the Milky Way halo contribution to the FRB sight-line, which is poorly constrained by current observations \citep{keating_2020_mnras}, to $\mathrm{DM_{Halo}^{MW} = DM - DM_{ISM}^{MW} - DM_{IGM} - DM_{M81}} \lesssim 32$--$42$\dmunit.
This is well in line with the models of \citet{yamasaki_2020_apj}, which predict  $\mathrm{DM^{MW}_{ Halo}}=10$--$30$\,\dmunit, but lower than predicted in some other models\citep{prochaska_2019_mnras}. 

Similar considerations are also applicable for the rotation measure\citep{bhardwaj_2021_apjl} $\mathrm{RM=-36.9\,rad\,m^{-2}}$ as determined in the companion paper by Nimmo et al. (submitted). The Galactic contribution\citep{hutschenreuter_2021_arxiv} along the line of sight to \frb\ is $\mathrm{RM=-17 \pm 4\,rad\,m^{-2}}$ (Figure~\ref{fig:dm_rm_map}). The contribution of the IGM is most likely minor, leaving only the MW halo, the M81 halo and the local environment of \frb as contributing sources. The RMs of the MW- and M81-halo are likely small ($\mathrm{|RM|<20\,rad\,m^{-2}}$), constraining $\mathrm{RM_{Local}}$ to the range $\mathrm{[+20, -60]\,rad\,m^{-2}}$. This is comparable to earlier results\citep{chime_2019_apjl} for \rthree ($\mathrm{RM=-115\,rad\,m^{-2}}$) but three orders of magnitude lower than the (highly variable) RM of \rone \citep{michilli_2018_natur}. This indicates that models of FRB sources do not necessarily require extreme magneto-ionic environments, unless the magnetic fields along the line of sight to both \frb\ and \rthree\ are strongly tangled, such that they result in a low net RM. More likely though, \rone\ resides in a very different environment giving rise to its observed properties.

\begin{table*}
\begin{center}
\caption{\label{tab:gc-properties}Notable properties of \gc.}
\begin{tabular}{lcc}
\hline
\hline
Property & Value & Reference\\
\hline
Metallicity $\mathrm{log}[\text{Z}/\text{Z}_{\odot}]$ & $-1.74^{0.80}_{-0.90}$ &  this work\\
Metallicity [Fe/H] & -1.83$^{0.86}_{-0.87}$ & this work\\
Stellar mass $\mathrm{log}[\rm{M/M_{\odot}}]$ & $5.77^{0.19}_{-0.22}$  & this work\\
Effective radius (R$_{\mathrm{eff}}$/pc) &  3.7 & this work\\
Age (Gyr) &  9.13$^{3.27}_{-4.18}$ & this work\\
$\mathrm{(u-r)_{0}^{a}}$ (AB mag)& 1.96(2) & \cite{alam_2015_apjs} \\
E(V$-$B)$\mathrm{^{b}}$ & 0.2$^{0.1}_{-0.1}$ & this work \\
$\sigma_{\rm r}$ (km\,s$^{-1}$) & 22  & this work \\
Absolute r-band mag. (AB) & $-$8.4 & -- \\
Luminosity distance (Mpc) & 3.6 & \cite{karachentsev_2005_aj} \\ 
 \hline
 \multicolumn{3}{l}{$\mathrm{^{b}}$ Milky Way extinction is corrected using a} \\
 \multicolumn{3}{l}{\hspace{0.3cm}reddening map \cite{sfd1998}.} \\
 \multicolumn{3}{l}{$\mathrm{^{c}}$ Using R$_{v}$ = 3.1.} \\
\end{tabular}
\end{center}
\end{table*}

\begin{table*}[ht]
\begin{center}
\hspace{-1.in}
\caption{Broadband SDSS filters used to model the SED of \gc. }
\label{filter:maggies}
\begin{tabular}{@{} *4l @{}}
\hline
\hline
 &  & Effective & Flux density\\
Instrument& Filter & Wavelength [\AA] & [maggie]$\mathrm{^{a}}$ \\ \midrule 

 SDSS & u & 3546 & 2.92$\times 10^{-9}$ \\
 & $g$ & 4670 & 1.05 $\times 10^{-8}$ \\
 & $r$ & 6156 & 1.77$\times 10^{-8}$ \\
 & $i$ & 7472 & 2.04$\times 10^{-8}$ \\
 & $z$ & 8917 & 2.33$\times 10^{-8}$\\ 
 \hline
 \multicolumn{4}{l}{$\mathrm{^{a}}$ The flux densities are assigned a 20\% fractional }\\
 \multicolumn{3}{l}{\hspace{0.3cm}uncertainty. Note that 1 maggie is defined }\\
 \multicolumn{3}{l}{\hspace{0.3cm}as the flux density in Jansky divided by 3631.}
\end{tabular}
\end{center}
\end{table*}

\begin{table*}[h]
\begin{center}
\caption{ Free parameters and their associated priors for the Prospector `delayed tau' model.}
\label{tab:sfhmodel}
\hspace{-1.in}
\begin{tabular}{lll}
 \hline
 Parameter & Description & Prior\\
 \hline
 log(M/M$_{\odot}$)   &  total stellar mass formed & uniform: min=3, max=7\\
 log(Z/Z$_{\odot}$)  &   stellar metallicity   & top-hat: min=-3.5, max=0 \\
 dust2 & diffuse V-band dust optical depth  &  top-hat: min=0.0, max=2.0\\
 t$\rm{_{age}}$   & age of \gc & top-hat: min=0.1, max=13.8\\
 $\tau$ & e-folding time of the SFH  & uniform: min=0.1, max=30\\
 \hline
\end{tabular}
\end{center}
\end{table*}

\section*{Data availability}
The datasets generated from the EVN observations and analysed in this study are available at the Public EVN Data Archive under the experiment codes EK048B, EK048C, and EK048F.

\section*{Code availability}
The codes used to analyse the data are available at the following sites:
    AIPS (\url{http://www.aips.nrao.edu/index.shtml}),
    CASA (\url{https://casa.nrao.edu}),
    Difmap (\url{ftp://ftp.astro.caltech.edu/pub/difmap/difmap.html}),
    DSPSR (\url{http://dspsr.sourceforge.net/}),
    FETCH (\url{http://dspsr.sourceforge.net/}),
    Heimdall (\url{https://sourceforge.net/projects/heimdall-astro/}),
    IRAF (\url{http://ast.noao.edu/data/software}),
    PRESTO (\url{https://github.com/scottransom/presto}),
    PSRCHIVE (\url{http://psrchive.sourceforge.net}), and 
    SpS (\url{https://github.com/danielemichilli/SpS}).

\bibliography{bib-entries}

\paragraph{Acknowledgements}
We would like to thank the directors and staff at the various participating stations for allowing us to use their facilities and running the observations. 
The European VLBI Network is a joint facility of independent European, African, Asian, and North American radio astronomy institutes. Scientific results from data presented in this publication are derived from the following EVN project code: EK048.
This work was also based on simultaneous EVN and PSRIX data recording observations with the 100-m telescope of the MPIfR (Max-Planck-Institut f\"{u}r Radioastronomie) at Effelsberg, and we thank the local staff for this arrangement.
We would like to express our gratitude to Willem van Straten for modifying the DSPSR software package to fit our needs. 
We appreciate helpful discussions about magnetar formation scenarios in a globular cluster environment with E. P. J. van den Heuvel.
A.B.P is a McGill Space Institute (MSI) Fellow and a Fonds de Recherche du Quebec - Nature et Technologies (FRQNT) postdoctoral fellow.
B.M. acknowledges support from the Spanish Ministerio de Econom\'ia y Competitividad (MINECO) under grant AYA2016-76012-C3-1-P and from the Spanish Ministerio de Ciencia e Innovaci\'on under grants PID2019-105510GB-C31 and CEX2019-000918-M of ICCUB (Unidad de Excelencia ``Mar\'ia de Maeztu'' 2020-2023).
Basic research in radio astronomy at NRL is funded by 6.1 Base funding. Construction and installation of VLITE was supported by the NRL Sustainment Restoration and Maintenance fund.
C.J.L. acknowledges support from the National Science Foundation Grant 2022546.
C.L. was supported by the U.S. Department of Defense (DoD) through the National Defense Science \& Engineering Graduate Fellowship (NDSEG) Program.
D.M. is a Banting Fellow.
E.P. acknowledges funding from an NWO Veni Fellowship.
F.K. acknowledges support from the Swedish Research Council via Grant No. 2014-05713.
FRB research at UBC is supported by an NSERC Discovery Grant and by the Canadian Institute for Advanced Research.
J.P.Y. is supported by the National Program on Key Research and Development Project (2017YFA0402602).
K.S. is supported by the NSF Graduate Research Fellowship Program.
K.W.M. is supported by an NSF Grant (2008031).
M.B. is supported by an FRQNT Doctoral Research Award.
N.W. acknowledges support from the National Natural Science Foundation of China (Grant 12041304 and 11873080)
P.S. is a Dunlap Fellow and an NSERC Postdoctoral Fellow. The Dunlap Institute is funded through an endowment established by the David Dunlap family and the University of Toronto. 
Part of this research was carried out at the Jet Propulsion Laboratory, California Institute of Technology, under a contract with the National Aeronautics and Space Administration. The NANOGrav project receives support from National Science Foundation (NSF) Physics Frontiers Center award number 1430284.
The Dunlap Institute is funded through an endowment established by the David Dunlap family and the University of Toronto. B.M.G. acknowledges the support of the Natural Sciences and Engineering Research Council of Canada (NSERC) through grant RGPIN-2015-05948, and of the Canada Research Chairs program.
The National Radio Astronomy Observatory is a facility of the National Science Foundation operated under cooperative agreement by Associated Universities, Inc. SMR is a CIFAR Fellow and is supported by the NSF Physics Frontiers Center award 1430284.
This work is based in part on observations carried out using the 32-m radio telescope operated by the Institute of Astronomy of the Nicolaus Copernicus University in Toru\'n (Poland) and supported by a Polish Ministry of Science and Higher Education SpUB grant.
V.B. acknowledges support from the Engineering Research Institute Ventspils International Radio Astronomy Centre (VIRAC).
V.M.K. holds the Lorne Trottier Chair in Astrophysics \& Cosmology and a Distinguished James McGill Professorship and receives support from an NSERC Discovery Grant and Herzberg Award, from an R. Howard Webster Foundation Fellowship from the Canadian Institute for Advanced Research (CIFAR), and from the FRQNT Centre de Recherche en Astrophysique du Quebec
We receive support from Ontario Research Fund—research Excellence Program (ORF-RE), Natural Sciences and Engineering Research Council of Canada (NSERC) [funding reference number RGPIN-2019-067, CRD 523638-201, 555585-20], Canadian Institute for Advanced Research (CIFAR), Canadian Foundation for Innovation (CFI), the National Science Foundation of China (Grants No. 11929301), Simons Foundation, Thoth Technology Inc, and Alexander von Humboldt Foundation. Computations were performed on the SOSCIP Consortium’s [Blue Gene/Q, Cloud Data Analytics, Agile and/or Large Memory System] computing platform(s). SOSCIP is funded by the Federal Economic Development Agency of Southern Ontario, the Province of Ontario, IBM Canada Ltd., Ontario Centres of Excellence, Mitacs and 15 Ontario academic member institutions.
Based [in part] on data collected at Subaru Telescope, which is operated by the National Astronomical Observatory of Japan.
The National Radio Astronomy Observatory (NRAO) is a facility of the National Science Foundation operated under cooperative agreement by Associated Universities, Inc.
{\sc AIPS} is a software package produced and maintained by the National Radio Astronomy Observatory (NRAO).

\paragraph{Author contributions}
F.K. is the Principal Investigator of the PRECISE team, he organised the observations, found the five bursts in the raw voltages and coordinated writing of the manuscript. B.M. lead the analysis of the correlated data, performed the localisation and wrote parts of the manuscript. K.N. lead the time-domain analysis of the bursts. J.W.T.H. lead the interpretation of the results and wrote parts of the manuscript. M.B. lead and performed the modelling of \gc. S.P.T. performed the data reduction and analysis of the Subaru data. A.K. wrote and modified the software correlator SFXC to allow for the highest time resolution data. J.Y. assisted with the data reduction, analysis and interpretation of the correlated data. M.P.S. helped with the manuscript and created Fig. \ref{fig:dm_rm_map}. P.S. and A.B.P. reduced and analyzed the archival Chandra data. C.J.L. reduced and analysed the Realfast data. W.M.P. reduced and analysed the VLITE data. M.G. searched the Fermi catalogues. Z.P. assisted with the reduction and analysis of the correlated data. C.B. assessed the optical registration errors of the Subaru images. D.M.H. searched the PSRIX and DFB data for bursts. U.B. coordinated and performed the observations at Effelsberg. V.B. coordinated and performed the observations at Irbene. M.B. helped commissioning the dual recording mode at SRT. S.T.B coordinated and performed the observations at Noto. J.E.C. supported the observations at Onsala. A.C. implemented the dual recording mode at SRT and performed some of the observations. R.F. supports the observations at Toru\'{n}. O.F. wrote observing schedules. M.P.G. coordinated and performed the observations at Toru\'{n}. R.K. assisted with the dual recording at Effelsberg. M.A.K. supports the observations at Badary, Svetloe and Zelenchukskaya. M.L. supported the observations at Onsala and assisted with the manuscript. G.M. coordinated and performed the observations at Medicina. A.M. coordinated and performed the observations at Badary, Svetloe and Zelenchukskaya. O.S.O-B wrote observing schedules. A.P. supported the observations at SRT. G.S. ran most of the observations at SRT. N.W. and J.Y. coordinated and performed the observations at Urumqi. V.M.K. played a significant coordination role that enabled these results.  All other co-authors contributed to the CHIME/FRB discovery of the source or the interpretation of the analysis results and the final version of the manuscript.

\paragraph{Competing interests}
The authors declare no competing interests.

\paragraph{Additional information}
\paragraph{Correspondence and requests for materials} should be addressed to F.K.

\end{document}